**Toward a Human–AI Task Tensor: A Taxonomy for Organizing Work in the Age of Generative AI**

Anil R. Doshi
Strategy and Entrepreneurship, UCL School of Management, anil.doshi@ucl.ac.uk

Alastair P. Moore
Operations and Technology, UCL School of Management, a.p.moore@ucl.ac.uk

**Abstract**
We introduce a framework for understanding the impact of generative AI on human work, which we call the human–AI task tensor. A tensor is a structured framework that organizes tasks along multiple interdependent dimensions. Our human–AI task tensor introduces a systematic approach to studying how humans and AI interact to perform tasks, and has eight dimensions: task definition, AI integration, interaction modality, audit requirement, output definition, decision-making authority, AI structure, and human persona. After describing the eight dimensions of the tensor, we provide illustrative frameworks (derived from projections of the tensor) and a human–AI task canvas that provide analytical tractability and practical insight for organizational decision-making. We demonstrate how the human–AI task tensor can be used to organize emerging and future research on generative AI. We propose that the human–AI task tensor offers a starting point for understanding how work will be performed with the emergence of generative AI.

**Takeaways**

- We introduce the Human–AI Task Tensor, a framework that organizes AI-enabled tasks along eight dimensions: task definition, AI integration, interaction modality, audit requirement, output definition, decision-making authority, AI structure, and human persona.
- By mapping tasks along the dimensions in the Human–AI Task Canvas, we propose that organizations and researchers can identify patterns that go beyond individual use cases, uncovering broader applications that guide effective human–AI collaboration across different domains and activities.
- The AI Function Matrix draws on the AI integration (as a complement or substitute) and the task's output definition (as well-defined or ill-defined) to organize six distinct functions an AI can play in a task: production, idea generation, assistance, editing, explanation, and open-ended interaction.
- The Task Augmentation/Automation Scale outlines 20 levels of decision-making authority in human–AI dyads to illustrate degrees of augmentation and automation.
- The Task Audit Matrix classifies human–AI interactions based on the need for process and output auditing, identifying four key task types: open exchange, verifiable application, process exploration, and expert application.

## 1. Introduction

In a short span of time, generative artificial intelligence (AI)[1] has emerged to play a significant role in human work and activities (Liang, et al., 2025). From an applied perspective, generative AI has distinct characteristics from prior breakthrough general-purpose technologies (Eloundou et al., 2023): inputs into the model are unstructured and require no technical or coding knowledge, both input and output can be multi-modal (e.g., numeric, text, audio, image, video) and non-deterministic; and it has distinctive anthropomorphic, sentient qualities (Chen et al., 2024; Cohn et al., 2024; Ibrahim et al., 2025; Shanahan, 2024). These qualities of generative AI suggest there are numerous possibilities for its use across many human activities inside and outside of organizations.

Indeed, the adoption of generative AI has proceeded at an unprecedented pace (Humlum & Vestergaard, 2024). At the same time, organizations, academics, and policymakers have engaged in a wave of research and analysis on its adoption, impacts, and implications. The motivation for much of these early studies has been to understand the role of generative AI in creating and improving outputs and making activities more efficient. Several outcomes (across multiple activities) have been studied, including productivity (Peng et al., 2023; Jia et al., 2024), creativity and problem solving (Boussioux et al., 2024; Doshi and Hauser, 2024), and scientific idea generation (Chai et al., 2024, Gottweis et al., 2024). The impact of generative AI on management and strategy has similarly emerged as an area of study, with initial work looking at how generative AI makes strategic evaluations (Csaszar et al., 2024; Doshi et al., 2024).

---

[1] Generative AI refers to AI systems designed to produce new content based on user input. Many current generative AI tools are large language models (LLMs), which are trained on vast datasets of text, enabling them to understand and generate human-like language. One example of a technology driving LLMs is the generative pre-trained transformer (GPT) architecture, which relies on a neural network structure called a transformer to process and generate sequential data (Vaswani et al., 2017). These models work by predicting the most probable next word (or token) in a sequence, which allows them to generate coherent and contextually relevant outputs.

Yet excitement about generative AI has been accompanied by an undercurrent of apprehension. This new technology is improving rapidly (Rein et al., 2024) and has uncertain impacts and capabilities that ostensibly compete with a wide range of human economic activities. The widespread availability of generative AI has raised concerns about job security (Felten, Raj, and Seamans, 2023), disincentivization of investment in human capital (Cowgill et al., 2024), and workplace demands (Wang et al., 2024).

Our goal in this chapter is to impose some structure around the complex, multi-dimensional problem of understanding tasks in the presence of human–AI partnerships by providing an initial taxonomy. By framing the emerging understanding of generative AI in the context of performing tasks, our taxonomy may be of use to both researchers seeking to understand the heterogeneous effects of generative AI and to practitioners seeking to implement generative AI in their organizations. The taxonomy has eight dimensions: task definition, AI integration, interaction modality, audit requirement, output definition, decision-making authority, AI structure, and human persona. Together, these dimensions can be aggregated into a human–AI task tensor that acts as an initial framework for organizing how a human and AI dyad might accomplish tasks. The term *tensor,* which we borrow from mathematics and computer science, is defined as a multidimensional representation of a system and transformation between states in the system. In our context, the tensor helps to conceptualize how tasks evolve within an eight-dimensional AI-influenced environment, which enables us to reduce complexity into analytically manageable frameworks.

We first provide a description of each of the dimensions of the AI task space. Next, we describe how different projections of the human–AI task tensor result in tractable and helpful frameworks for assessing tasks and performance in the presence of generative AI. We present the tensor as a visual "canvas" to help map tasks across the eight dimensions. Finally, we show how the tensor can be used to organize and partly reconcile research findings, and provide a roadmap for future research. We conclude with some potential considerations that may affect how the tensor evolves.

## 2. Eight human–AI task tensor dimensions

We delineate tasks performed by a human–AI dyad on eight dimensions. The dimensions can be organized around three broad stages of a task: its formulation, implementation, and resolution. Formulation includes outlining the scope of and parameters of the task; implementation involves execution and result generation; and resolution relates to task evaluation and the transmission of results to the appropriate recipient.

Table 1 summarizes the eight dimensions in the human–AI task tensor. Our assignment of dimensions to the three stages of a task corresponds primarily to when they are performed versus when they are planned or decided. Some dimensions may overlap across stages. Moreover, our categorization is not exhaustive but rather should serve as an instructive guideline for mapping each dimension to the time horizon of task execution.

[INSERT TABLE 1 ABOUT HERE]

### 2.1. Task definition

The first dimension in the task tensor pertains to how well-defined a task is. Task definition calls back to classic research on how well structured a problem is. Simon (1973) distinguishes well- and ill-defined problems along several characteristics pertaining to whether criteria exist for testing solutions and whether a state space can represent the problem and solutions. Similarly, Rittel and Webber (1973) describe "wicked problems" as those that are not well formulated and for which not all relevant information can be known when stating the problem. We consider an analogue to these characteristics for tasks. Well-defined tasks have more complete information, clearer criteria for evaluation, and a more easily defined sub-task space. Some tasks are well-defined for either the human or AI, such as sorting a list. Other tasks that lack those traits may be more ill-defined, such as developing a representation of a business strategy (Csaszar, 2018).

The extent to which a task is well- or ill-defined might affect how it is performed by the human or AI, or shared between the two, and, by extension, how preferable AI augmentation or automation is. A well-defined task might be easily performed by AI if that task is within the AI's capabilities (Dell'Acqua et al., 2023); in such cases, ensuring technical feasibility is likely important (Kwa et al., 2025). For ill-defined tasks, human judgment or perspective may create greater diversity in outcomes, which may be more or less desirable depending on the context.

### 2.2. AI integration

The second dimension, AI integration, represents how the AI is being used in the dyad to accomplish the task vis-à-vis the human's effort. Specifically, the AI might be incorporated into the performance of a task by either substituting for or complementing the human's effort, including their skills, bundles of skills, or job. An index that assesses the risk of job substitution by AI can be created by evaluating the capabilities of generative AI against the work required in different jobs across the economy (Felten et al., 2023). Early labor-market studies have demonstrated the threat of substitution by showing fewer job posts and freelancer employment and lower income following the introduction of generative AI tools (Demirci et al., 2025; Hui et al., 2024). AI may also act as a complement to humans in a particular workflow—for example, by promoting previously distant skills as being more relevant and complementary to accomplish a specific task (Krakowski et al., 2022; Mäkelä and Stephany, 2025).

We consider the possibility of AI acting as a substitute or complement at the task level. For many tasks, the open-ended nature of inputs and outputs into generative AI allow it to substitute human work by producing it instead or to complement human work by enhancing or improving it in some way. While incentives might drive substitution in some areas, generative AI can be used either to substitute or to complement.

### 2.3. Interaction modality

The interaction modality of a task can be either digital or physical, with each having significant implications for AI's capabilities. Today's AI models primarily operate in digital contexts, receiving and generating structured or unstructured information such as numeric data, text, images, audio, and video.

To interact with AI systems, analog inputs must first be converted into digital form. However, developments in embodied AI—such as robotics, haptics, kinetics, and object manipulation—are progressively extending AI's reach into the physical world. When integrated into physical systems (e.g., autonomous robots, augmented reality interfaces, or voice-activated assistants), AI becomes embodied, gaining the ability to directly sense, understand, and manipulate real-world objects and environments. This embodiment represents a transformative expansion that allows AI not only to process and produce digital information but also to physically act within and respond dynamically to its real-world context. Thus, the extent to which the modality is digital or physical will affect which tasks are undertaken and how they are performed.

### 2.4. Audit requirement

The fourth dimension is the audit requirement, which spans the implementation and resolution stages. In a human–AI collaboration, one of the agents will be performing work (i.e., the provider agent) that will be passed to the other agent (i.e., the recipient agent). The work might be an intermediate output that the recipient agent would use for subsequent work, or it might be the task's final output. The audit requirement refers to the level of oversight required from the recipient agent of both the process underpinning the work produced and the output.[2]

Regarding the process, one key question is: How much knowledge about the process used to perform the work does the recipient agent require to assess the work? Some processes may operate in a "black box," such that the recipient agent simply needs to receive the output and does not need to know the process that led to it. In other cases, by contrast, knowledge and review of the underlying process may be important to either the human or the AI. For example, in the event of systematic failure by AI (Shneiderman, 2020), it may be socially desirable to maintain human knowledge redundancies with machines. Conversely, an AI audit of the process may assist with ongoing improvements in model training and fine tuning.

Regarding the output, it is important to know the extent to which the recipient agent requires the ability to audit the output provided by the provider agent. Some exchanges do not require audit or verification. For example, asking for an opinion or additional thoughts on some subject merely contributes to the dialogue or exchange. There is no need to confirm whether those outputs were optimal or correct, per se. Other types of interactions may yield outputs that need to be verified. When the output is factual or strategic in nature, for example, the recipient agent may need to evaluate one or more outputs provided by the provider agent to make an assessment.

[2] Our notion of audit resembles the idea of "scalable oversight" by Amodei et al. (2016), who describe the problem of providing feedback for training automated systems when the task is complex or the objective function is only partly defined.

An example that demonstrates both process and output audit is asking an AI system to develop software code to run an economic model, such as a market-clearing mechanism. An AI likely would be able to take instructions about the task as input and produce as output its assumptions and the resulting code, which may successfully run without any errors. Yet a process audit would be desirable to ensure that the underlying assumptions about the mechanism were properly coded, and an output audit would be desirable to ensure the result is consistent with expectations.

**2.5. Output definition**

Output definition covers the extent to which the output that completes the task is well- or ill-defined (and is thus a corollary to task definition). While generative AI is capable of producing either type of output (assuming they are both in a medium that the AI model is capable of producing), there may be differences in how the task is performed by the human–AI dyad and how it is evaluated. A well-defined output may have some outcome of utility, quality, value, or effectiveness associated with it and thus may be more easily audited or evaluated. Well-defined outputs, such as numerical results in a financial report, can be precisely measured or evaluated against objective criteria. Ill-defined outputs, such as a mind map, often require human interpretation and refinement. Their ambiguous nature may make it difficult to assess whether the output is comprehensive and whether it contains errors of omission or commission.

**2.6. Decision-making authority**

Decision-making authority is characterized by one of the two agents in the dyad holding decision-making rights for the task (i.e., the decision-maker) and the other serving as the associate to the decision-maker (i.e., the contributor). Decision-making authority captures the degree of augmentation versus automation of the task (Raisch & Krakowski, 2021). When the human is the decision-maker, AI can provide varying degrees of augmentation, with its role increasing as augmentation increases. When the AI is the decision-maker, greater involvement by the human contributor represents decreased automation.

### 2.7. AI structure

Wood (1986) argues that when assessing task completion, the task itself should be distinguished from the performer of the task. As a result, the final two dimensions encompass the two agents in the dyad, the AI and the human. Both dimensions are present in all the three task stages.

The structure of an AI system can be understood through the arrangement of its components as it matures along a spectrum from genesis (early, experimental forms) toward utility (standardized, mature forms). At the genesis end of the spectrum, AI systems typically feature novel architectures or configurations, involving highly customized or exploratory arrangements; novel and specialized models; and experimental data flows designed for specific tasks or contexts. Over time, these structures tend to move toward utility, becoming increasingly standardized, modular, and robust, with well-defined roles and predictable interactions between components or agents (Arthur, 2007). Importantly, as initial models move toward being increasingly standardized, new approaches to, say, data, modeling, or implementation may mark new instances of novel architectures.

Management choices shift from bespoke architectures to standardized modules, clearly defined interfaces, and commonly accepted data-handling practices. This evolution enhances reliability, interoperability, and ease of integration with broader technological ecosystems. Research indicates that modular architectures in AI systems can lead to better generalization, scalability, and interpretability, and can apply to both models and data (Sun & Guyon, 2023).

The key question for managers is thus to determine the degree to which AI components should be standardized into reliable utilities versus retaining custom, exploratory structures possibly optimized for specific or novel tasks.

### 2.8. Human persona

Finally, the human is central to the task. Different types of people will respond to and be impacted by generative AI in different ways. Research has shown that the impact of generative AI in different contexts varies based on characteristics of the user. In some contexts, the benefits of AI accrue disproportionately to those at the lower end of the distribution of ability, measured in different ways (Brynjolfsson et al., 2023; Dell'Acqua et al., 2023; Doshi & Hauser, 2024; Peng et al., 2023). In other contexts, those who are most capable benefit disproportionately (Otis et al., 2023; Roldán-Monés, 2024). There are undoubtedly several human characteristics relevant to understanding performance in the human–AI dyad, including ability, expertise, and experience.

## 3. Illustrative projections of the human–AI task tensor

The space described by the eight dimensions of the human–AI task tensor captures different aspects of a task that affect how it is implemented and how well human–AI dyads complete it. Each dimension is measured on a continuous scale and is thus difficult to use for conceptualization. To assist with this effort, we provide a few illustrative projections of the human–AI task tensor into more simplified spaces, or frameworks, that can offer theoretical tractability or practical insights for managers.

### 3.1. The AI Function Matrix (based on AI integration and output definition)

Several longstanding and more recent typologies are available for categorizing types of work and tasks. Considering the extent to which a problem is analyzable and has prominent exceptions, Perrow (1967) identifies four types of tasks: routine (analyzable, few exceptions), engineering (analyzable, many exceptions), craft work (unanalyzable, few exceptions), and nonroutine (unanalyzable, many exceptions). These types loosely map onto Davenport's (2005) categories of tasks, which are based on the level of complexity and collaboration required: transactional, contribution, expert work, and collaboration. Hackman and Oldham (1975) consider five dimensions of a job: the variety of skills required, the clarity of the outcome, its significance, the level of required autonomy, and feedback (while completing the task and from others).

For generative AI, work tasks have been considered as use cases for the technology. Mollick and Mollick (2023) describe seven task areas for using generative AI in education: feedback, instruction, metacognition, collaborative intelligence, AI as student, and practice opportunities. Zhao-Sanders (2024) lists one hundred specific tasks that people use generative AI for, including writing and editing documents (including scripts, emails, and code) and translating text or code.

Our conception of AI functions completed by a human–AI dyad is based on the following two dimensions: AI integration and output definition. Across these two dimensions, we define an AI Function Matrix in which we identify six types of AI functions in performing a task: production, idea generation, assistance, editing, explanation, and open-ended. Table 2 presents the AI Function Matrix.

[INSERT TABLE 2 ABOUT HERE]

AI can largely substitute for human effort by producing output or generating ideas. Output might be well defined, such as an information-retrieval or fact-finding task, or less defined, such as a mind map outlining an academic literature or the text of an email. When it comes to idea generation, AI might perform the initial locus of work, rather than fully producing the output. Ideas can be generated for well- or ill-defined tasks. For example, an innovation context is well-defined (where the ideas can be measured by how novel and useful they are; Harvey & Berry, 2022), while suggesting activities to do in a new city is ill-defined (where ideas might also depend on preferences).

AI might also complement work begun by humans. AI could be set up to provide assistance, such as providing critiques of writing or suggesting additions to a report. AI could also be used to review human output, such as copyediting text or changing the tone of an email. AI might further complement humans in tasks that lack a defined outcome. AI can offer explanations or guidance, such as helping a student learn a new topic or a lawyer understand the terms of a new contract. Alternatively, because generative AI's output is open-ended—language, in the case of LLMs—AI can engage in open exchange, such as acting as a companion or advisor that a human might turn to for help, comfort, or companionship (Yin et al., 2024). In these cases, the AI might complement human perception and understanding.

### 3.2. Task Augmentation/Automation Scale (based on degrees of decision-making authority)

To provide additional insight into the decision-making authority dimension, we delineate 20 discrete types of authority, based on the degree of automation or augmentation that exists between the decision-maker and the contributor agents. The 20 types include ten with a human decision-maker and ten with an AI decision-maker (with each set of ten effectively reflecting the other). Table 3 shows the 20 types in the Task Augmentation/Automation Scale. When the human is the decision-maker (left-hand side of the table), the involvement of the AI contributor decreases down the table, indicating decreasing levels of augmentation. When the AI is the decision-maker (right-hand side), the involvement of the human contributor decreases, indicating increasing levels of automation.

[INSERT TABLE 3 ABOUT HERE]

In the first three rows, the contributor(s) support the decision-maker by providing options and assistance. Human decision makers might consider the AI as providing low, medium, or high decision support. Similar to the functioning of a company board or the combination of models in a machine learning ensemble, the value from a multiplicity of decision inputs depends on how expert or specialized the decision input is and how correlated the errors are between different contributors (Doshi et al., 2024). The following five rows represent increased autonomy of the decision-maker with ever-decreasing intervention of the contributor. From most to least interventionist, the contributor can play the following roles: giving approval ("in the loop;" Wu et al., 2022), holding veto authority ("on the loop;" Nahavandi, 2017), observing the decision ("near the loop;" Jackson & Pinto, 2024), having the option to request the decision ("aware of the loop"), and being available to be informed at the decision-maker's discretion ("invited to the loop;" Johnson et al., 2011). Each of these interventions represents possible types of involvement by the contributor, without having input into the decision. The final two levels represent full autonomy for the decision-maker.

Row 9 represents decisions where information is lost because the contributor was not involved in or informed of the decision.[3] If tacit knowledge is not codified (Ancori et al., 2000) and decisions shared with AI are a codification of that tacit knowledge, then decisions not shared by a human with AI remain tacit and outside the purview of an organization's AI capabilities.

[3] For the human decision-maker, we labeled this "SNAFU" to represent the possibility of intentional or unintentional human concealment of the decision. For the AI decision-maker, we labeled this "WOPR" in reference to the 1983 film *WarGames*, in which the computer, WOPR, elected to initiate a weapon launch sequence without informing humans.

When AI is unaware of a decision, the AI cannot use that data for training, reinforcement, or other learning activities. Conversely, one concern about AI is the possibility it will make level 9 decisions in which humans should have some involvement (as described by automation levels 1 to 8) but do not. Some in the AI community are concerned that AI will act not only independently of humans, but without human knowledge (Armstrong et al., 2013; Center for AI Safety, 2023). The final row represents full automation (when the AI makes the decision) and non-automation (when the decision is left solely to the human).

The augmentation side of the table overlaps with Parasuraman et al.'s (2000) levels of automation. We extend their model by allowing for either the human or the AI to act as the decision-maker. When breaking down the level of intervention by the contributor to either the human or AI decision-maker, the Task Augmentation/Automation Scale shows not only levels of automation but also levels of augmentation. Augmentation and automation are essentially the converse of one another. The assumption underlying augmentation and automation is that the human owns the decision rights, and machines enter to assist (i.e., to augment) or take over a task (i.e., automate). When, instead, either the human or the AI can hold decision rights, the relationship between the two can be considered more symmetrically. In other words, moving down the Task Augmentation/Automation Scale is associated with greater autonomy to the decision-maker and less involvement by the contributor. Along the lines of Raisch and Krakowski's (2021) argument that the delineation between augmentation and automation is often too stark, the Task Augmentation/Automation Scale provides a stepwise view of both of these processes that shows many possible gradations of how human work is being augmented or automated.

### 3.3. Task Audit Matrix (based on process and output audit)

In Table 4, our Task Audit Matrix summarizes four types of interactions between human and AI based on whether the process or outcome requires auditing.

[INSERT TABLE 4 ABOUT HERE]

When neither requires an audit, both the human and the AI engage in an open interaction of free exchanges without concern for error or failure. For example, casual conversations do not require an audit. Conversely, interactions that require an audit of both the process and the outcome are expert applications, where expert knowledge is required to interpret the process by which an outcome was realized as well as the outcome itself. For example, when the AI is used for data analysis, the human likely needs some understanding of statistics and coding to audit how computations were made and to understand whether the final outputs are valid or correct. The type of interaction that results when the process does not require an audit but the outcome does is a verifiable application. In this case, the human may cede certain knowledge or expertise, since the output is sufficient to assess the validity of the work performed by the AI. For example, a human may need to evaluate ideas that AI provides for a writing task, but the human may not necessarily need to understand the creative process underlying those ideas (and, indeed, likely does not even when the task is performed by another human). When the process requires an audit, but the outcome does not, we consider these to be process explorations, where an understanding of the drivers or mechanisms is critical. For example, when a human is learning how to negotiate, the process by which an AI advisor ostensibly takes a position is important, but the position itself may not have a need or way to be audited.

## 4. The Human–AI Task Canvas

The illustrated frameworks above show various ways in which the tensor can be projected to help organize human–AI tasks. In Figure 1, we also present the tensor as a visual graphic, the Human-AI Task Canvas, that can be used to depict how tasks might be performed with the involvement of AI. The canvas provides a structured way to assess where a given task falls along all dimensions of the task tensor (unlike the reduced frameworks). The canvas reflects other similar tools from strategy, including the strategy canvas (Kim & Mauborgne, 2015), and from technology, such as the I-space (Boisot & Cox, 1999). Like those tools, the canvas provides a way to visualize individual tasks and compare multiple tasks across all eight dimensions.

[INSERT FIGURE 1 ABOUT HERE]

Each dimension is represented on a continuous scale. The scales for both task definition and output definition run from ill-defined to well-defined. AI integration is on a scale from complement to substitute, to reflect the possibility that the AI might be integrated in a way that is not purely one or the other. Interaction modality accounts for the extent to which the inputs or outputs are digital or physical. The audit dimension separately accounts for the required levels of both process and output audit. Decision-making authority is determined by whether the human or AI holds decision-making authority. If the human is the decision-maker, the scale reflects the degree to which AI augments their decisions. If AI is the decision-maker, the scale reflects the extent to which it automates human input. The AI structure is represented by the level of customization required. The human persona dimension captures whether the human user is a "low-type" or "high-type" on the relevant characteristic.

### 4.1. Organizing current research and mapping future research

We can use the tensor and accompanying canvas to assess different use cases in practice and in research and to understand results on generative AI that are emerging across research disciplines. It can also be used to understand areas that might be fruitful for future research on generative AI and work.

The human–AI task tensor and canvas can be used to illustrate how the same task can be approached differently by a human–AI dyad. For example, early research looked at how generative AI affected workplace activities, namely in call centers (Brynjolfsson et al., 2023; Jia et al., 2024). In Jia et al. (2024), an AI voice agent was given the job of making outbound calls to assess interest in a financial product. The work task in this case could be described using select human–AI task tensor dimensions as: AI integration as substitute (doing the work of having the conversation), high AI decision-rights autonomy (no human intervention occurred in the lead-generation phase), or low AI audit (the human took the output of leads from the AI without verification). Also in call centers, Brynjolfsson et al. (2023) studied a different region in the human–AI task tensor. Specifically, they looked at a work task that had more complementary AI integration (suggestions for responses and related support documentation), lower AI decision-rights autonomy (the human elected to follow the AI's suggestions), and a higher audit requirement (the human can assess which, if any, suggestion is most appropriate). We propose that describing human–AI work using the human–AI task tensor will allow for clearer differentiation of the work tasks studied.

The human–AI task tensor might also reconcile results in different contexts. For example, research by Doshi and Hauser (2024) and Otis et al. (2024) suggests a possible difference in groups that disproportionately benefit from generative AI. Doshi and Hauser (2024) show that improvement in creative story-writing from generative AI ideas accrue disproportionately among the least creative individuals. By contrast, Otis et al. (2024) show that advice to entrepreneurs from generative AI largely benefits the top-performing entrepreneurs. Though characteristics of the human users in these studies are not directly comparable (i.e., inherent creativity versus entrepreneurial ability, respectively), these results point in different directions to what part of the distribution benefits most from access to generative AI. In Doshi and Hauser (2024), the AI output required relatively little auditing by the human (the writer could use or adapt the ideas as they preferred, and ideas did not require verification or oversight), the task was well-defined (write a story), and the AI structure included the use of a general LLM. By comparison, Otis et al. (2024) required a high degree of auditing (the entrepreneur likely benefits from understanding the process driving the suggestions and the validity of the suggestions), the task was relatively less defined (improve entrepreneurial performance), and the AI structure included specific instructions to fit the responses more closely to the task. One plausible explanation for the difference in results, then, would arise from how each is represented in the tensor on these dimensions. Human users in both cases used the AI to generate ideas (for stories and business strategies, respectively). However, the definition of the tasks was different. Also, in a task with relatively little need for auditing, users were free to incorporate ideas as they liked, and the ideas kickstarted performance among those who were less inherently capable. In a task with relatively greater need for auditing, users who possessed the training or knowledge to conduct the audit derived greater value from the AI outputs.

Finally, the dimensions of the tensor can also help identify new areas for future research. For instance, research has largely focused on LLMs and textual interactions, with some work showing their impact in different media (Zhou & Lee, 2024). Future research could investigate how generative AI affects human or aggregate output across different media, such as video and audio, when used more as a substitute or more as a complement, especially as generative AI's capabilities develop in those areas. Other research might vary who the decision-maker is (human or AI) and the extent of the contributor's involvement, and assess outcomes and how the human engages with the task. A third possibility might be to explore how generative AI performs when varying task and output definition. Finally, the tensor itself may also be studied to improve its completeness. Studying different tasks at scale might unearth additional or refined dimensions to further improve on our initial effort at crafting the framework. In general, the tensor can assist researchers with mapping existing research and identifying gaps to find new avenues of inquiry.

## 5. Discussion

Generative AI is poised to have a significant impact on management and organizations. This impact promises to be multi-faceted and applicable across an organization, its processes and design, and its products and services. The atomic unit in which humans and AI jointly work is the task (Vaccaro et al., 2024). We have proposed a human–AI task tensor that can assist with structuring tasks performed jointly by a human–AI dyad. The tensor consists of eight dimensions: task definition, AI integration, interaction modality, audit requirement, output definition, decision-making authority, AI structure, and human persona. Our hope is that the human–AI task tensor will enhance understanding of organizational design and the allocation of tasks within organizations in the presence of AI (Puranam, 2021) and AI system design within organizations (Akata et al., 2020). We also compiled the eight dimensions into a Human–AI Task Canvas, which allows researchers and managers to visualize and compare different tasks.

Different dimensions of the human–AI task tensor might provide guidance on how to use generative AI, given individuals' needs beyond accomplishing a task. For example, the audit and AI integration dimensions pertain to questions about human knowledge acquisition and skill formation. The ability to audit requires the human to have preexisting knowledge of and skills in the task being performed. When an individual lacks the requisite and skills, using AI as a substitute hampers the effort required to develop that knowledge and those skills (Dell'Acqua, 2024; Wiles et al., 2024). One normative implication, then, is that generative AI might serve as an aid in human knowledge and skill development when used as a complement (with appropriate intention and interaction) and as a hindrance when used as a substitute. Thus, decisions of when and how to apply generative AI may also have important implications for how humans specialize with the availability of generative AI.

The human–AI task tensor also has practical implications for managers. Organizations are at the initial stages of grappling with the large and multi-faceted question of how to integrate generative AI into their processes (Davenport & Alavi, 2023; Davenport & Mittal, 2022; Puntoni et al., 2023; Ramge & Mayer-Schönberger, 2023; Roslansky, 2023; Wilson & Daugherty, 2024). The human–AI task tensor offers a schema for organizing the implementation of generative AI based on where the task resides in the tensor, rather than by function or specific use cases. By observing dimensions of the human–AI task tensor where the organization is experiencing greater value and success, managers might be able to apply successful applications of generative AI to tasks in nearby regions. Specifically, as managers experiment with AI applications and use cases, they will gather information on areas where a value proposition is more clearly realized. For those tasks, if managers can identify other tasks that are nearby in the tensor (i.e., those that are similar along most, or all, dimensions), then they can more quickly deploy existing AI applications to those tasks, and thus may realize greater or more immediate gains. Tasks that are distant in the tensor may require a different solution.

Notably, our human–AI task tensor is not the first effort at imposing a framework on how humans and AI might collectively perform work or tasks (Choudhary et al., 2023; Dellermann et al., 2021; Parasuraman et al., 2000). We believe our approach complements existing frameworks because, with our focus on downstream tasks currently performed by humans, we take an applied view from the perspective of the organization and organizational design. Moreover, the human–AI task tensor may be incomplete. For example, other dimensions might reasonably be considered relevant to the tensor. Nonetheless, we believe our human–AI task tensor offers a useful way to classify tasks to help organize theoretical implications and practical applications of generative AI. The tensor can help researchers organize fieldwork and managers deploy AI in their organizations.

How generative AI evolves will also affect the human–AI task tensor. We partly account for the evolving nature of generative AI's models, applications, and methods with our AI structure and interaction modality dimensions. However, we may not have captured all of the implications that will accompany generative AI's evolution. Modes of interaction or implementation of future incarnations of AI may alter how we define work tasks, what a human–AI collaboration looks like, and how more complex strategies are broken down into discrete tasks. For example, what are the implications of an AI that is capable of a "super automation" of sorts—when it can devise the problem, break the problem into constituent component tasks, and perform those tasks? Moreover, the possibility of artificial general intelligence raises its own set of questions that extends well beyond work tasks (Morris et al., 2024).

Our human–AI task tensor also does not explicitly consider the ethics of generative AI involvement in different tasks or the implications of that involvement for (human) social welfare. There are certainly instances where task allocation may be disproportionately allocated to humans, even in situations where AI can perform them capably. For example, humans individually or collectively may not want to lose the competency, skills, or knowledge required to perform certain tasks. Alternatively, there may be value in humans performing tasks to facilitate connection with oneself and with others. Questions concerning the "optimal" allocation of tasks between humans and AI are critical but likely cannot be answered by the human–AI task tensor alone.

As generative AI emerges in the coming years, tasks within organizations will change and adapt to the presence of artificial workers. How managers plan for these changes will depend partly on the types of tasks that are performed throughout the organization. As these changes take place, researchers will concurrently be conducting further investigations on the impact and implications of generative AI for organizations. Our human–AI task tensor provides a structure that can be used in industry and academia to organize these changes. If humans and organizations are on the verge of experiencing one of the biggest redistributions in how work is performed between humans and technology, then the human–AI task tensor may serve as a map to navigate the transition.

## References

Amodei, D., Olah, C., Steinhardt, J., Christiano, P., Schulman, J., & Mané, D. (2016). Concrete problems in AI safety. *arXiv*. https://doi.org/10.48550/arXiv.1606.06565

Ancori, B., Bureth, A., & Cohendet, P. (2000). The economics of knowledge: The debate about codification and tacit knowledge. *Industrial and Corporate Change, 9*(2), 255–287. http://doi.org/10.1093/icc/9.2.255

Akata, Z. et al. (2020) A Research agenda for hybrid intelligence: Augmenting human intellect with collaborative, adaptive, responsible, and explainable artificial intelligence. *Computer, 53*(8), 18–28. doi.org/10.1109/MC.2020.2996587

Armstrong, S., Bostrom, N., & Shulman, C. (2013). Racing to the precipice: a model of artificial intelligence development. *AI & Society, 31*, 201–206. doi.org/10.1007/s00146-015-0590-y

Arthur, W. B. (2007). The Structure of Invention, *Research Policy*, *36*(2), 274–287. https://doi.org/10.1016/j.respol.2006.11.005

Boisot, M., & Cox, B. (1999). The I-Space: a framework for analyzing the evolution of social computing, *Technovation, 19*(9), 525-536. https://doi.org/10.1016/S0166-4972(99)00049-8

Boussioux, L., Lane, J. N., Zhang, M., Jacimovic, V., Lakhani, K. R. (2024). The crowdless future? Generative AI and creative problem-solving. *Organization Science, 35*(5), 1589–1607. https://doi.org/10.1287/orsc.2023.18430

Brynjolfsson, E., Li, D., & Raymond, L. R. (2025). Generative AI at work. *The Quarterly Journal of Economics*, qjae044. https://doi.org/10.1093/qje/qjae044

Center for AI Safety. (2023). Statement on AI risk. https://www.safe.ai/statement-on-ai-risk

Chai, S., Doshi, A. R., & Tröbinger, M. (2025). Generative AI in scientific ideation: How research experience moderates perception and integration. *SSRN*. https://dx.doi.org/10.2139/ssrn.5013086

Cheng, M., DeVrio, A., Egede, L., Blodgett, S. L., & Olteanu, A. (2024). "I am the one and only, your cyber BFF": Understanding the impact of GenAI requires understanding the impact of anthropomorphic AI. *arXiv*. https://doi.org/10.48550/arXiv.2410.08526

Choudhary, V., Marchetti, A., Shrestha, Y. R., & Puranam, P. (2025). Human-AI ensembles: When can they work? *Journal of Management, 51*(2), 536–569. https://doi.org/10.1177/01492063231194968

Cohn, M., Pushkarna, M., Olanubi, G. O., Moran, J. M., Padgett, D., Mengesha, Z., & Heldreth, C. (2024). Believing anthropomorphism: Examining the role of anthropomorphic cues on trust in large language models. *Extended Abstracts of the 2024 CHI Conference on Human Factors in Computing Systems, Association for Computing Machinery, New York, NY, USA, 54*, 1–15. https://doi.org/10.1145/3613905.3650818

Cowgill, B., Hernandez-Lagos, P., & Wright, N. L. (2024). Does AI cheapen talk? Theory and evidence from global entrepreneurship and hiring. *SSRN*. https://dx.doi.org/10.2139/ssrn.4702114

Csaszar, F. A. (2018). What makes a decision strategic? Strategic representations. *Strategy Science, 3*(4), 606–619. https://doi.org/10.1287/stsc.2018.0067

Csaszar, F. A., Ketkar, H., & Kim, H. (2024) Artificial intelligence and strategic decision-making: Evidence from entrepreneurs and investors. *Strategy Science, 9*(4), 322–345. https://doi.org/10.1287/stsc.2024.0190

Davenport, T. H. (2005). *Thinking for a living: How to get better performance and results from knowledge workers*. Harvard Business School Press.

Davenport, T. H., & Mittal, N. (2022, November 14). How generative AI is changing creative work. *Harvard Business Review*. https://hbr.org/2022/11/how-generative-ai-is-changing-creative-work

Davenport, T., & Alavi, M. (2023, July 6). How to train generative AI using your company's data. *Harvard Business Review*. https://hbr.org/2023/07/how-to-train-generative-ai-using-your-companys-data

Dell'Acqua, F. (2024). Falling asleep at the wheel: Human/AI collaboration in a field experiment on HR recruiters. Working paper.

Dell'Acqua F., McFowland, E., Mollick, E., Lifshitz-Assaf, H., Kellogg, K. C., Rajendran, S., Krayer, L., Candelon, F., Lakhani, K. R. (2023) Navigating the jagged technological frontier: Field experimental evidence of the effects of AI on knowledge worker productivity and quality. *SSRN*. https://dx.doi.org/10.2139/ssrn.4573321

Dellermann, D., Calma, A., Lipusch, N., Weber, T., Weigel, S., & Ebel, P. (2021). The future of human-AI collaboration: A taxonomy of design knowledge for hybrid intelligence systems. *arXiv*. https://doi.org/10.48550/arXiv.2105.03354

Demirci, O., Hannane, H., Zhu, X. (2025) Who is AI Replacing? The impact of generative AI on online freelancing platforms. *Management Science*. https://doi.org/10.1287/mnsc.2024.05420

Doshi, A. R., Bell, J. J., Mirzayev, E., & Vanneste, B. S. (2025). Generative artificial intelligence and evaluating strategic decisions. *Strategic Management Journal, 46*, 583–610. https://doi.org/10.1002/smj.3677

Doshi, A. R., & Hauser, O. P. (2024). Generative artificial intelligence enhances individual creativity but reduces the collective diversity of novel content. *Science Advances, 10*(28), eadn5290. https://doi.org/10.1126/sciadv.adn5290

Eloundou, T., Manning, S., Mishkin, P., & Rock, D. (2024). GPTs are GPTs: An early look at the labor market impact potential of large language models. *Science, 384*, 1306–1308. https://doi.org/10.1126/science.adj0998

Felten, E. W., Raj, M., & Seamans, R. (2023). Occupational heterogeneity in exposure to generative AI. *SSRN*. https://dx.doi.org/10.2139/ssrn.4414065

Gottweis, J., Weng, W. H., Daryin, A., Tu, T., Palepu, A., Sirkovic, P., Myaskovsky, A., Weissenberger, F., Rong, K., Tanno, R., Saab, K., Popovici, D., Blum, J., Zhang, F., Chou, K., Hassidim, A., Gokturk, B., Vahdat, A., Kohli, P., Matias, Y., Carroll, A., Kulkarni, K., Tomasev, N., Dhillon, V., Vaishnav, E. D., Lee, B., Costa, T. R. D., Penadés, J. R., Peltz, G., Xu, Y., Pawlosky, A., Karthikesalingam, A., & Natarajan, V. (2025). Towards an AI co-scientist. *arXiv*. https://doi.org/10.48550/arXiv.2502.18864

Hackman, J. R., & Oldham, G. R. (1975). Development of the job diagnostic survey. *Journal of Applied Psychology, 60*(2), 159–170. https://doi.org/10.1037/h0076546

Harvey, S., & Berry, J. (2022). Toward a meta-theory of creativity forms: How novelty and usefulness shape creativity. *Academy of Management Review, 48*(3), 504–529. https://doi.org/10.5465/amr.2020.0110

Hui, X., Reshef, O., Zhou, L. (2024) The short-term effects of generative artificial intelligence on employment: Evidence from an online labor market. *Organization Science* 35(6):1977-1989. https://doi.org/10.1287/orsc.2023.18441

Humlum, A., & Vestergaard, E. (2024). The unequal adoption of ChatGPT exacerbates existing inequalities among workers. *Proceedings of the National Academy of Sciences, 122*(1), e2414972121. https://doi.org/10.1073/pnas.2414972121

Ibrahim, L., Akbulut, C., Elasmar, R., Rastogi, C., Kahng, M., Morris, M. R., McKee, K. R., Rieser, V., Shanahan, M., & Weidinger, L. (2025). Multi-turn evaluation of anthropomorphic behaviours in large language models. *arXiv*. https://doi.org/10.48550/arXiv.2502.07077

Jackson, J. M., Pinto, M.D. (2024). Human near the loop: Implications for artificial intelligence in healthcare. *Clinical Nursing Research, 33*(2-3), 135–137. https://doi.org/10.1177/10547738241227699

Jia, N., Luo, X., Fang, Z., & Liao, C. (2024). When and how artificial intelligence augments employee creativity. *Academy of Management Journal, 67*(1), 5–32. https://doi.org/10.5465/amj.2022.0426

Johnson, M., Bradshaw, J.M., Feltovich, P.J., Jonker, C.M., van Riemsdijk, B., Sierhuis, M. (2011). The fundamental principle of coactive design: Interdependence must shape autonomy. In: De Vos, M., Fornara, N., Pitt, J.V., Vouros, G. (eds) *Coordination, Organizations, Institutions, and Norms in Agent Systems VI. COIN 2010. Lecture Notes in Computer Science(), Berlin, Heidelberg, 6541*. https://doi.org/10.1007/978-3-642-21268-0_10

Kim, W. C., & Mauborgne, R. A. (2015). *Blue Ocean Strategy: How to Create Uncontested Market Space and Make the Competition Irrelevant*. Harvard Business Review Press.

Krakowski, S., Luger, J., Raisch, S. (2022). Artificial intelligence and the changing source of competitive advantage, *Strategic Management Journal, 44*(6), 1425–1452. https://doi.org/10.1002/smj.3387

Kwa, T., West, B., Becker, J., Deng, A., Garcia, K., Hasin, M., Jawhar, S., Kinniment, M., Rush, N., Von Arx, S., Bloom, R., Broadley, T., Du, H., Goodrich, B., Jurkovic, N., Miles, L. H., Nix, S., Lin, T., Parikh, N., Rein, D., Sato, L. J. K., Wijk, H., Ziegler, D. M., Barnes, E., & Chan, L. (2024). Measuring AI ability to complete long tasks. arXiv. https://arxiv.org/abs/2503.14499

Liang, W., Zhang, Y., Codreanu, M., Wang, J., Hancheng, C., & Zou, J. (2025). The Widespread adoption of large language model-assited writing across society. *arXiv*. https://doi.org/10.48550/arXiv.2502.09747

Mäkelä, E., & Stephany, F. (2024) Complement or substitute? How AI increases the demand for human skills. *arXiv*. https://doi.org/10.48550/arXiv.2412.19754

Morris, M. R., Sohl-Dickstein, J., Fiedel, N., Warkentin, T., Dafoe, A., Faust, A., Farabet, C., & Legg, S. (2024). Levels of AGI for operationalizing progress on the path to AGI. *Proceedings of the 41st International Conference on Machine Learning (ICML), Vienna, Austria*. https://openreview.net/forum?id=0ofzEysK2D

Mollick, E., & Mollick, L. (2023). Assigning AI: Seven approaches for students with prompts. *SSRN*. https://dx.doi.org/10.2139/ssrn.4475995

Nahavandi, S. (2017). Trusted autonomy between humans and robots: Toward human-on-the-loop in robotics and autonomous systems. *IEEE Systems, Man, and Cybernetics Magazine, 3*(1), 10–17. https://doi.org/10.1109/MSMC.2016.2623867

Otis N., Clarke R., Delecourt S., Holtz D., Koning R. (2024) The uneven impact of generative AI on entrepreneurial performance. Working paper.

Parasuraman, R., Sheridan, T. B., & Wickens, C. D. (2000). A model for types and levels of human interaction with automation. *IEEE Transactions on Systems, Man, and Cybernetics - Part A: Systems and Humans, 30*(3), 286–297. https://doi.org/10.1109/3468.844354

Park, J. S., Morris, M. R., O'Brien, J. C., Liang, P., Cai, C. J., & Bernstein, M. S. (2023). Generative agents: Interactive simulacra of human behavior. *Proceedings of the 36th Annual ACM Symposium on User Interface Software and Technology (UIST '23), San Francisco, CA, USA. ACM*. https://doi.org/10.1145/3586183.3606763

Peng, S., Kalliamvakou, E., Cihon, P., & Demirer, M. (2023). The impact of AI on developer productivity: Evidence from GitHub Copilot. *arXiv*. https://doi.org/10.48550/arXiv.2302.06590

Perrow, C. (1967). A framework for the comparative analysis of organizations. *American Sociological Review, 32*(2), 194–208. https://doi.org/10.2307/2091811

Puntoni, S., Ensing, M., & Bowers, J. (2024. May 24). How marketers can adapt to LLM-powered search. *Harvard Business Review*. https://hbr.org/2024/05/how-marketers-can-adapt-to-llm-powered-search

Puranam P. (2021) Human–AI collaborative decision-making as an organization design problem. *Journal of Organizational Design, 10*(2):75–80. https://doi.org/10.1007/s41469-021-00095-2

Qiu, R., Xu, Z., Bao, W., & Tong, H. (2024). Ask, and it shall be given: Turing completeness of prompting. *arXiv*. https://doi.org/10.48550/arXiv.2411.01992

Raisch, S., & Krakowski, S. (2021). Artificial intelligence and management: The automation–augmentation paradox. *Academy of Management Review*, 46(1), 192–210. https://doi.org/10.5465/amr.2018.0072

Ramge, T., & Mayer-Schönberger, V. (2023, August 24). Using ChatGPT to make better decisions. *Harvard Business Review*. https://hbr.org/2023/08/using-chatgpt-to-make-better-decisions

Rein, D., Hou, B. L., Cooper Stickland, A., Petty, J., Pang, R. Y., Dirani, J., Michael, J., & Bowman, S. R. (2024). GPQA: A graduate-level Google-proof Q&A benchmark. *arXiv*. https://doi.org/10.48550/arXiv.2311.12022

Rittel, H. W. J., & Webber, M. M. (1973). Dilemmas in a general theory of planning. *Policy Sciences, 4*, 155–169. https://doi.org/10.1007/BF01405730

Roldán-Monés, A. (2024). When GenAI increases inequality: evidence from a university debating competition. Working paper.

Roslansky, R. (2023, December 4). Talent management in the age of AI. *Harvard Business Review*. https://hbr.org/2023/12/talent-management-in-the-age-of-ai

Shanahan, M. (2024). Talking about large language models. *Communications of the ACM, 67*(2), 68–79. https://doi.org/10.1145/362472

Shneiderman, B. (2020). Human-Centered Artificial Intelligence: Reliable, Safe & Trustworthy. *International Journal of Human–Computer Interaction, 36*(6), 495–504. https://doi.org/10.1080/10447318.2020.1741118

Simon, H. A. (1973). The structure of ill-structured problems. *Artificial Intelligence, 4*(3–4), 181–201. https://doi.org/10.1016/0004-3702(73)90011-8

Sun, H., & Guyon, I. (2023). Modularity in deep learning: A survey. In: Arai, K. (eds) Intelligent Computing. SAI 2023. Lecture Notes in Networks and Systems, 739. https://doi.org/10.1007/978-3-031-37963-5_40

Vaccaro, M., Almaatouq, A. & Malone, T. (2024). When combinations of humans and AI are useful: A systematic review and meta-analysis. *Nature Human Behaviour, 8*, 2293–2303. https://doi.org/10.1038/s41562-024-02024-1

Vaswani, A., Shazeer, N., Parmar, N., Uszkoreit, J., Jones, L., Gomez, A. N., Kaiser, Ł., & Polosukhin, I. (2017). Attention is all you need. In *Proceedings of the 31st Conference on Neural Information Processing Systems (NIPS 2017)*.

Wang, J., Doshi, A. R., & Landis, B. Experimental Evidence on the Within-Person Effects of Using Generative Artificial Intelligence. Working paper.

Wiles, E., Krayer, L., Abbadi, M., Awasthi, U., Kennedy, R., Mishkin, P., Sack, D., & Candelon, F. (2024). GenAI as an exoskeleton: Experimental evidence on knowledge workers using GenAI on new skills. *SSRN*. https://dx.doi.org/10.2139/ssrn.4944588

Wilson, H. J., & Daugherty, P. R. (2024). Embracing Gen AI at work. *Harvard Business Review*. https://hbr.org/2024/09/embracing-gen-ai-at-work

Wood, R. (1986). Task complexity: Definition of the construct, *Organizational Behavior and Human Decision Processes, 37*(1), 60–82. https://doi.org/10.1016/0749-5978(86)90044-0
Wu, X., Xiao, L., Sun, Y., Zhang, J., Ma, T., & He, L. (2022). A survey of human-in-the-loop for machine learning. *Future Generation Computer Systems, 135*, 364–381. https://doi.org/10.1016/j.future.2022.05.014
Yin, Y., Jia, N., & Wakslak, C. J. (2024). AI can help people feel heard, but an AI label diminishes this impact. *Proceedings of the National Academy of Science, 121*(14), e2319112121. https://doi.org/10.1073/pnas.231911212
Zao-Sanders, M. (2024, March 19). How people are really using GenAI. Harvard Business Review. https://hbr.org/2024/03/how-people-are-really-using-genai.
Zhou, E., & Lee, D. (2024). Generative artificial intelligence, human creativity, and art. *PNAS Nexus, 3*(3), pgae052. https://doi.org/10.1093/pnasnexus/pgae052
Zhou, P., Pujara, J., Ren, X., Chen, X., Cheng, H. T., Le, Q. V., Chi, E., Zhou, D., Mishra, S., & Zheng, H. (2024). SELF-DISCOVER: Large language models self-compose reasoning structures. *Advances in Neural Information Processing Systems*, 37, 126032–126058. https://proceedings.neurips.cc/paper_files/paper/2024/hash/e41efb03e20ca3c231940a3c6917ef6f-Abstract-Conference.html

## Tables and Figures

Table 1. Summary of human–AI task tensor dimensions

| Dimension | Description | Task Stage | | |
|---|---|---|---|---|
| | | Formulation | Implementation | Resolution |
| Task definition | The extent to which the task is ill-defined or well-defined | ✓ | | |
| AI integration | The extent to which the AI acts as a substitute or complement to the human | | ✓ | |
| Interaction modality | Whether the format of output is digital or physical | | ✓ | |
| Audit requirement | How much oversight of the process and output is required | | ✓ | ✓ |
| Output definition | The extent to which the output is ill defined or well defined | | | ✓ |
| Decision-making authority | Whether the AI or human is the decision-maker and the involvement of the contributor | | | ✓ |
| AI structure | How the AI is set up to accomplish a task (i.e., its data, modeling approach, customization, and implementation) | ✓ | ✓ | ✓ |
| Human persona | Type of human user in the task (e.g., whether the human engaging in the task is a low-type or high-type) | ✓ | ✓ | ✓ |

Table 2. The AI Function Matrix

| | | **Output definition** | |
|---|---|---|---|
| | | **Well defined** | **Less defined** |
| **AI integration** | **Substitute** | Production<br>Idea generation | |
| | **Complement** | Assistance<br>Editing | Explanation<br>Open |

Table 3. Augmentation/automation Scale

| | Human with decision rights | | AI with decision rights | | |
|---|---|---|---|---|---|
| **Augmentation level** | **Role of AI contributor** | **Colloquial analogy** | **Role of human contributor** | **Colloquial analogy** | **Automation level** |
| 10 | Many AI-aided options | Low-level decision support | Many human-aided options | Wisdom of the crowds | 1 |
| 9 | Few AI-aided options | Medium-level decision support | Few human-aided options | Few-shot learning | 2 |
| 8 | Single AI-aided option | High-level decision support | Single human-aided option | One-shot learning | 3 |
| 7 | AI approval | Model in the loop | Human approval | Human in the loop | 4 |
| 6 | AI veto power | Model on the loop | Human veto power | Human on the loop | 5 |
| 5 | Human informs machine | Model near the loop | AI informs human | Human near the loop | 6 |
| 4 | Human informs machine upon request | Model aware of the loop | AI informs human upon request | Human aware of the loop | 7 |
| 3 | Human informs machine if it decides to | Model invited to the loop | AI informs human if it decides to | Human invited to the loop | 8 |
| 2 | Human does not inform the machine | SNAFU | AI does not inform human | WOPR | 9 |
| 1 | No AI assistance | Human executed | No human assistance | AI automated | 10 |

Table 4. Task Audit Matrix

| | | **Output audit requirement** | |
|---|---|---|---|
| | | **Low** | **High** |
| **Process audit requirement** | **Low** | Open exchange | Verifiable application |
| | **High** | Process explorations | Expert application |

Figure 1. The Human–AI Task Canvas

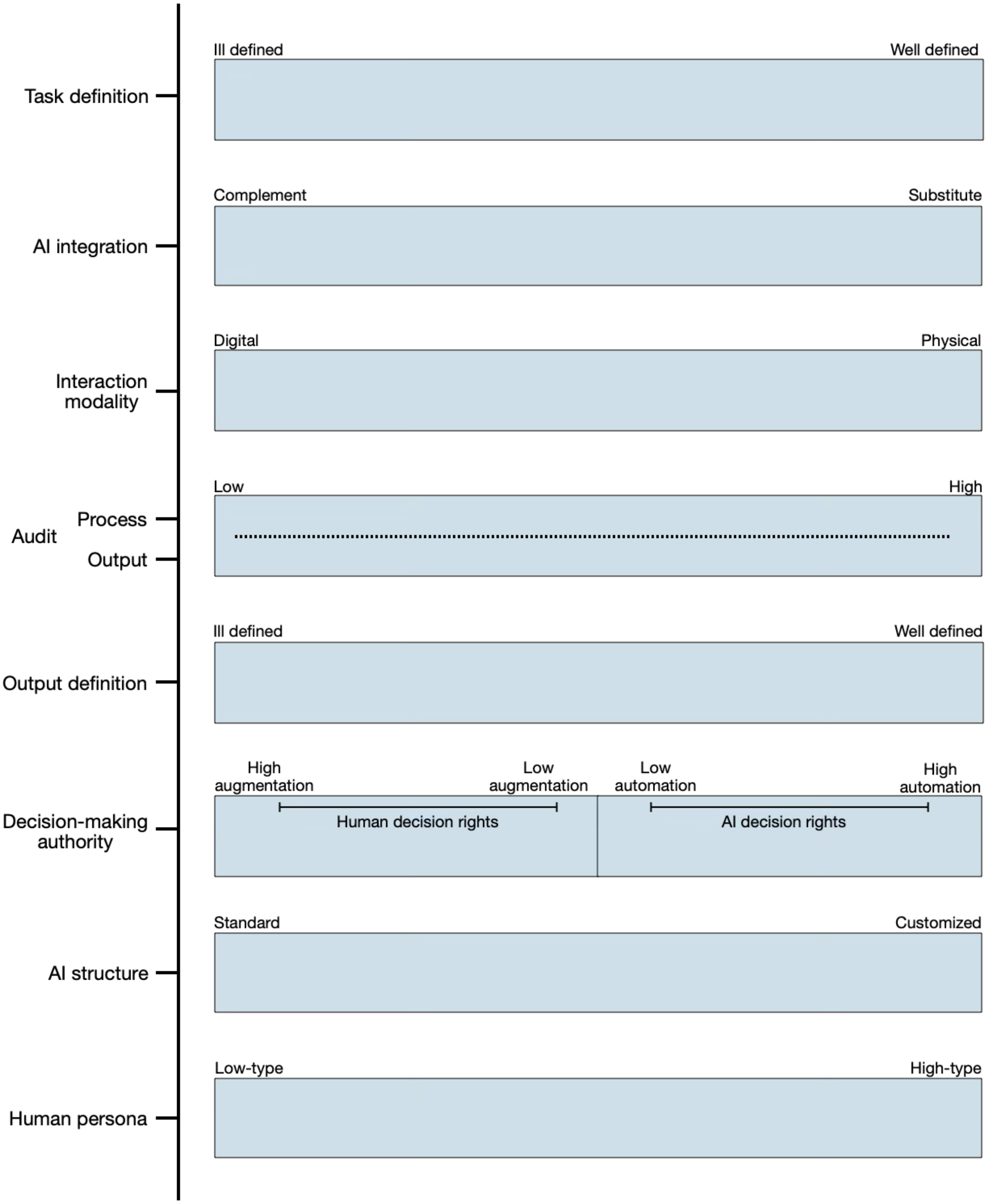

## Alt-Text

### Table 1: Summary of human–AI task tensor dimensions

Alt-text: A table summarizing the eight dimensions of the human–AI task tensor, grouped into three task stages: formulation, implementation, and resolution. The table lists each dimension—task definition, AI contribution, interaction modality, audit requirement, output definition, decision-making authority, AI structure, and human persona—and indicates which task stage(s) it applies to with checkmarks.

### Table 2: The AI Function Matrix

Alt-text: A matrix that categorizes AI task types based on two dimensions: AI contribution (substituting or complementing human effort) and output definition (well-defined or less defined). The table organizes six task types: production, idea generation, explanation, assistance, editing, and open-ended, mapping them to their respective positions in the framework.

### Table 3: Augmentation/automation Scale

Alt-text: A structured table displaying ten levels of decision-making authority in human–AI dyads. It contrasts human and AI decision rights across increasing autonomy levels, providing descriptions and colloquial analogies for each level. Levels range from full human control with AI support to full AI autonomy with no human involvement.

**Table 4: Task Audit Matrix**

Alt-text: A 2×2 table classifying tasks based on whether process and/or output audits are required. The categories include open exchange, verifiable application, process explorations, and expert application.

**Figure 1. The Human–AI Task Canvas**

A conceptual chart representing eight dimensions of the human-AI task tensor. Each vertical bar depicts a continuous spectrum on which each of the eight dimensions can vary. Task definition varies from ill-defined to well-defined. AI integration varies from complement to substitute. Interaction modality varies from digital to physical. Process and output audit requirements vary from low to high. Output definition varies from ill-defined to well-defined. Decision-making authority is split by whether the human or AI has decision authority. AI structure varies from standard to customized. Human persona varies from low-type to high-type.

## Author Biographies

### Anil R. Doshi

Anil Doshi is a strategy professor at the UCL School of Management. In his research on generative AI, he studies how new data, information, and content produced by AI models affect different areas of human activity, including creativity, decision-making, knowledge creation, and work. Anil earned his doctorate from Harvard University.

### Alastair P. Moore

Alastair Moore is an entrepreneur, educator, and investor. He is currently on the faculty at UCL School of Management and is a co-founder of Satalia.com and Deepflow.com. He has a PhD in machine learning from UCL.